\documentstyle[twocolumn,aps,psfig]{revtex}

\begin{document}
\draft

\twocolumn[\hsize\textwidth\columnwidth\hsize\csname     
@twocolumnfalse\endcsname

\title{Strongly Anisotropic Electronic Transport at Landau Level 
Filling Factor $\nu = 9/2$ and $\nu = 5/2$ Under Tilted Magnetic Field}

\author{W. Pan$^{a,b}$, 
R.R. Du$^{c,b}$,
H.L. Stormer$^{d,e}$, D.C. Tsui$^a$, L.N. Pfeiffer$^d$, 
K.W. Baldwin$^d$, and K.W. West$^d$}
\address{
a) Dept. Electr. Eng., Princeton University, Princeton, NJ}
\address{
b) NHMFL, Tallahassee, FL}
\address{
c) Dept. Phys., University of Utah, Salt Lake City, UT}
\address{
d) Bell Labs, Lucent Technologies, Bell Laboratories, Murray Hill, NJ}
\address{
e) Dept. Phys. and Dept. Appl. Phys., Columbia University, New York, NY}
 
\date{\today}
\maketitle
 
\begin{abstract}
We have investigated the influence of an increasing in-plane magnetic
field on the states at half-filling of Landau levels ($\nu$ = 11/2, 9/2,
7/2, and 5/2) of a two-dimensional electron system. In the electrically
anisotropic phase at $\nu$ = 9/2 and 11/2 an 
in-plane magnetic field of $\sim$ 1-2 T
overcomes its initial pinning to the crystal lattice 
and {\it reorient} this phase.
In the initially isotropic phases at $\nu$ = 5/2 and 7/2 an in-plane magnetic
field {\it induces} a strong electrical anisotropy. In all cases, for high
in-plane fields, the high resistance axis is parallel to the direction
of the in-plane field.
\end{abstract}
\vskip2pc]

The electrical transport properties at half-filling of either spin state of
Landau levels of a two-dimensional
electron system (2DES) have turned out to be very diverse. In the lowest
Landau level  (filling factor $\nu < 2$) at
half-filling of the down-spin level ($\nu = 1/2$) and at half-filling of the
up-spin level ($\nu = 3/2$) the Hall resistance,
$R_H$, is linear in magnetic field, $B$, and the magneto resistance, $R$, is only
weakly temperature dependent \cite{jiang:prb89}. Today,
this behavior is interpreted as the formation of composite fermions (CFs)
that fill up a fermi sea to a fixed fermi
wave vector, $k_F$, and the cancellation of the externally applied magnetic
field \cite{heinonen:book98,sarma:book98}.

In the second Landau level ($4 < \nu < 2$) at half-filling of the down-spin
level ($\nu = 5/2$) and half filling of the
up-spin level ($\nu = 7/2$) the Hall resistance shows a plateau 
and the magneto
resistance exhibits the deep minimum
of the fractional quantum Hall effect (FQHE) \cite{willett:prl87}. 
The occurrence of a FQHE
state at such an {\it even}-denominator
filling is puzzling, since from simple symmetry requirements on the wave
function, one would have expected the
FQHE to occur only at {\it odd}-denominator filling \cite{haldane:inbook87}. 
The origin of this
even-denominator FQHE remains
mysterious, but is now conjectured to arise from the formation of CF pairs
that condense into a novel state \cite{more:np91,greiter:prl91,morf:prl98}.
 
In higher Landau levels ($\nu > 4$) at half-filling of either spin level
($\nu$ = 9/2, 11/2, 13/2, 15/2) the Hall resistance,
$R_H$, is erratic and $R$ exhibits a strongly anisotropic behavior, showing a
strong peak in one current direction
($R_{xx}$) and a deep minimum when the current 
direction is rotated by ~90$^{\circ}$
within the plane ($R_{yy}$) \cite{stormer:march93,lilly:prl99,du:ssc99}. The
origin of these states remain also unclear. They are believed to arise from
the formation of a striped electronic
phase \cite{koulakov:prl96,moessner:prb96} 
or an electronic phase akin to a liquid crystal phase
\cite{fradkin:cont9810148}. The
broken symmetry is speculated to arise
from a slight misalignment of the crystallographic direction of the
GaAs/AlGaAs host lattice, which pins the
phase \cite{lilly:prl99,du:ssc99}. The wealth of 
different behaviors makes the states at
half-filling presently one of the most
fascinating topics in 2D electron physics in a high magnetic field.

Tilting the magnetic field with respect to the sample normal is a
classical method to gently alter the
conditions for the 2DES \cite{fang:pr68}.
In an ideal 2DES a tilted magnetic field does
not modify the orbital motion but only
the Zeeman splitting of its spin level. In a real 2DES, which has a finite
thickness of $\sim$ 100\AA, the orbital motion is
affected only to second order. Measuring the influence of such a tilt on
the transport parameters often allows to
narrow down the range of possible underlying electronic states.

Such measurements have been performed extensively on the standard
odd-denominator FQHE states \cite{haug:prb87} as
well as on the half-filling states at 
$\nu$ = 1/2, 3/2, 5/2, and 7/2 \cite{syphers:ssc88,eisenstein:prl88}. The
interpretation of the data largely draws
from the increase of the Zeeman energy under tilt. Such experiments have
been instrumental in supporting and
expanding the CF model around $\nu$ = 1/2 and 3/2 
\cite{du:prl95,gee:prb96} and they suggest the
involvement of the spin-degree of
freedom in the formation of the states at 
$\nu$ = 5/2 and 7/2 \cite{eisenstein:prl88}. For higher
Landau levels such angular-dependent
measurements have not yet been performed. Moreover, the in-plane anisotropy
of the resistance in this filling
factor regime introduces a new variable into such tilt experiments. 

Previous tilt experiments in the regime of the FQHE always assumed
that only the angle of the magnetic field
with respect to the sample normal mattered to the transport behavior,
whereas the azimuth of the field, {\it i.e.} the
direction of the so-created in-plane magnetic field ($B_{ip}$), was immaterial.
While this implicit assumption
should always have been suspect, since the in-plain current direction
breaks the planar isotropy, it obviously
needs to be investigated and justified in the case of the $\nu$ = 9/2, 11/2
states, which show strongly anisotropic phases.
In its extremes, the $B$-field can be tilted towards the direction that shows
the maximum in $R$ ($R_{xx}$, hard
direction) and towards the direction that shows the minimum in $R$ ($R_{yy}$, easy
direction), which is rotated with
respect to $R_{xx}$ by approximately 90$^{\circ}$ 
within the plane of the sample.

We have performed such tilt experiments on the states at half-filling
and observed very different behavior for
different states. For the states at 
$\nu$ = 9/2 and 11/2 the initial direction of
the in-plane anisotropy is overwritten by
the in-plane field. Depending on the tilt direction, and therefore the
direction of the in-plane field, the easy
direction and the hard direction either remain in place or trade places
with increasing $B_{ip}$. More surprisingly yet,
the $\nu$ = 5/2 and 7/2 states, 
which {\it do not show} any initial in-plane anisotropy
become strongly anisotropic under
tilt, to a degree similar as the states at 
$\nu$ = 9/2 and $\nu$ = 11/2. In all cases,
under high tilt angles, it is exclusively the
relative direction of current ($I$) and 
in-plane magnetic field ($B_{ip}$), that
determines whether $R$ shows a minimum
or a maximum. More specifically, in the large in-plane limit,  $R$ is a deep
minimum when measured
{\it perpendicular} to the in-plane magnetic field and it exhibits a strong
maximum when measured {\it parallel} to the
in-plane field. At the same time, neither the half-filled states in the
lowest Landau levels ($\nu$ = 1/2 and 3/2) nor
any of the FQHE states in their vicinity show such anisotropies. The origin
of the effect that an in-plane
magnetic field exerts onto the states at half-filling remains unclear.

Our sample consists of a modulation-doped GaAs/AlGaAs heterostructure
which has a 2DES with an
electron density of $2.2 \times 10^{11}$ cm$^{-2}$ 
and a low-temperature mobility of
$\mu = 1.7 \times 10^7$ cm$^2$/V~sec. The specimen is
similar to the one used in our previous, untilted experiments
\cite{du:ssc99} on the
$\nu$ = 9/2 and 11/2 states, but has a yet higher
mobility. The size of the sample is about 
4 mm $\times$ 4 mm with eight indium
contacts placed symmetrically around
the edges, four at the sample corners and four in the center of the four
edges. The sample is placed on a precision
rotator inside the mixing chamber of a dilution refrigerator placed within
a superconducting magnet. The
equipment reaches a base temperature of 40 mK  in magnet fields up to $B$ =
18 T. The sample can be rotated
{\it in-situ} around an axis perpendicular to the field from 
$\theta=0^{\circ}$ to $\theta = 90^{\circ}$.
Experiments are performed at fixed angle
$\theta$ while sweeping $B$. Since for our 
sample of fixed electron density only the
magnetic field perpendicular to the
2DES ($B_{perp}$ = $B \times$ cos($\theta$)) 
determines the filling factor $\nu$, we plot our data
against $B_{perp}$. At any given angle $\theta$ the
in-plane field ($B_{ip}=B \times$ sin($\theta$)) 
is then proportional to the perpendicular
magnetic field, {\it i.e.} $B_{ip} = B_{perp} \times$ tan($\theta$).
With a total magnetic field of 18 T available and an electron density of $2.2
\times 10^{11}$ cm$^{-2}$, which requires a $B_{perp}$
of $\sim$ 3.6 T to reach the $\nu$ = 5/2 state, 
the sample can be tilted as much as $\theta$ =
arccos(3.6T/18T) $\approx 78^{\circ}$. This creates an
in-plane magnetic field as high as 
$B_{ip} = B_{perp} \times$ tan(78$^{\circ}$) $\approx$ 17 T.

The sample was mounted in two different configurations onto the
rotator. In the first instance, the axis of
rotation was along y, the easy direction (low resistance) of the 9/2 and
11/2 states, allowing to place increasing
$B_{ip}$ along the hard direction (high resistance) 
(see inserts top Fig.~2). In
the second instance, the axis of rotation
ran along x, the hard direction (high resistance) of the 9/2 and 11/2
states, allowing to place increasing $B_{ip}$ along
the easy direction (low resistance) (see inserts bottom Fig. 2). Since the
hard and easy direction are not very
precisely defined within the plane, but only known to run roughly along the
edges of the square sample, the
in-plane field in our experiment may not run precisely along either the
hard or the easy direction, but it will run
predominantly in such a direction. Transport experiments were performed
using standard 7 Hz look-in
techniques at a current of 5 nA which is known from previous experiments
\cite{du:ssc99}
on these samples to cause negligible
electron heating. The transport anisotropy was measured at 14 different
angles between $\theta = 0^{\circ}$ and $\theta = 78^{\circ}$ in both
configurations. The angle $\theta$ was determined from the 
orderly cos($\theta$) shift of
several strong minima of the FQHE.

Fig.~1 shows an overview over $R_{xx}$ and $R_{yy}$ at 
zero-tilt ($B_{perp} = B$,
$B_{ip}$ = 0). The well-documented strong
anisotropy of the $\nu$ = 9/2, 11/2, 13/2, and 15/2 states is apparent in the
data. States at filling factor $\nu < 4$ show
negligible anisotropy. Any residual difference between 
$R_{xx}$ and $R_{yy}$ in this
regime can be attributed to a slight
difference in the geometry of the contact arrangement for both
measurements. In the following tilt experiments
we will focus on the states at $\nu$ = 9/2 and 11/2 
as well as $\nu$ = 5/2 and 7/2. The
states at half-filling of the next higher
Landau level, $\nu$ = 13/2 and 15/2 show behavior similar to the $\nu$ = 9/2 and 11/2
states, although less well pronounced.

Fig.~2 shows $R_{xx}$ and $R_{yy}$ data for 
$6 > \nu > 2$ at selected tilt angles, $\theta$.
The data of the top panels (Fig.~2a,b) are
taken with $B_{ip}$ pointing along 
the {\it hard} direction, x, of the anisotropic
state, whereas the data of the bottom
panel (Fig.~2c,d) are taken for $B_{ip}$ along 
the {\it easy} direction, y, of the
anisotropic state. The inserts depicts the
geometries. The behavior of $R_{xx}$ and $R_{yy}$ 
as a function of tilt differs
dramatically between the upper and the
lower panels.

For $\nu$ = 9/2 and 11/2 in the absence of tilt 
($\theta = 0.0^{\circ}$), the traces for $R_{xx}$
(solid lines in Fig.~2a and Fig.~2c) and the
traces for $R_{yy}$ (dotted lines in Fig.~2a and Fig.~2c) are essentially
identical. This reinforces our assertion, that
cycling of the sample to room temperature, necessary to change the sample
configuration, has negligible effect
on the transport features. As the sample is 
tilted toward $\theta =74.3^{\circ}$ the $R_{xx}$
and the $R_{yy}$ traces behave very
differently in both panels. When 
$B_{ip}$ is increased along the {\it hard} direction
(Fig.~2a) $R_{xx}$ is somewhat reduced in
amplitude but recovers at the highest tilt angles while 
$R_{yy}$ lifts up only
slightly from its value at $\theta = 0^{\circ}$.
Nevertheless, the maximum remains a maximum and the minimum remains a
minimum. On the other hand,
when $B_{ip}$ is increased along the {\it easy} 
direction (Fig.~2c) $R_{xx}$ collapses and
develops into a minimum, while $R_{yy}$
rises and becomes a maximum at the highest tilt. Here maximum and minimum
{\it trade places}. At the highest
angles the shape of $R_{yy}$ in Fig.~2c practically equals 
$R_{xx}$ in Fig.~2a and
vice versa.

For $\nu$ = 5/2 and 7/2, in the absence of tilt, the data show
practically no anisotropy. For this $\theta = 0^{\circ}$ situation,
$R_{xx}$ and $R_{yy}$ are very similar within Fig.~2b and both are very similar
within Fig.~2d. However, tilting of the
sample and the associated increase of $B_{ip}$ drastically alters the data and
introduces a strong anisotropy between
$R_{xx}$ and $R_{yy}$. As $B_{ip}$ increases along the x-direction (Fig.~2b), $R_{xx}$
increases, while $R_{yy}$ decreases. On the other
hand, as $B_{ip}$ increases along the y-direction (Fig.~2d), $R_{xx}$ decreases,
while $R_{yy}$ increases. The hard direction
always develops {\it along} $B_{ip}$, whereas the easy direction always runs
{\it perpendicular} to $B_{ip}$. This means that, the
directionality of this anisotropy is determined by the direction of $B_{ip}$.
This is particularly apparent at the highest
angel shown, $\theta=74.3^{\circ}$, where $R_{xx}$ and $R_{yy}$ seem to have traded places when
going from Fig.~2b to Fig.~2d.
Furthermore, at such large angles the anisotropy in the $\nu$ = 5/2 and 7/2
states becomes similar to the anisotropy in
the $\nu$ = 9/2 and 11/2 states. Independent of the 
starting conditions at $\theta = 0^{\circ}$,
eventually the direction of $B_{ip}$ governs
the directionality of the anisotropy for all such states at $\nu$ = 11/2, 9/2,
7/2, and 5/2. On the other hand, no such
anisotropy --- neither preexisting nor induced --- is found at $\nu$ = 3/2 nor at
$\nu$ = 1/2 in the lowest Landau level in tilted
magnetic field (not shown). This observation seems to link the states at
$\nu$ = 11/2 and 9/2 with the states at $\nu$ = 7/2
and 5/2. Beyond the equivalence in their high-angle anisotropy, even the
general shape of $R_{xx}$ and $R_{yy}$
approach each other at such high angles.

Fig.~3 summarizes the anisotropies for the strongest of states at
$\nu$ = 9/2 and $\nu$ = 5/2. The four top panels (Fig.~3a,b,c,d)
match the four panels of Fig.~2a,b,c,d.  They show the amplitudes
of  $R_{xx}$ (solid line) and $R_{yy}$ (dotted
line) at $\nu$ = 9/2 and $\nu$ = 5/2 filling versus the strength of the in-plane
magnetic field. The bottom panels of Fig.~3
represent the anisotropies of the $\nu$ = 9/2 state and the $\nu$ = 5/2 state as
calculated from the panels above. 

For $\nu$ = 9/2, when $B_{ip}$ increases along the x-direction (Fig.~3a) the
amplitude of $R_{xx}$ parallel to $B_{ip}$, drops
rapidly by more than a factor of two, reaches a minimum strength at $B_{ip}$
$\sim$ 2 T
and then recovers at the highest
fields to about 70$\%$ of its original value. The amplitude of the $R_{yy}$,
perpendicular to $B_{ip}$, rises somewhat from
zero, reaches a shallow maximum also at $B_{ip} \sim$ 2 T, and decays slightly for
higher fields.

On the other hand, when $B_{ip}$ increases along the y-direction (Fig.~3c)
the amplitude of $R_{xx}$ perpendicular to
$B_{ip}$, collapses precipitously, almost 
touching zero at $B_{ip} \sim$ 1 T and remains at
about 5$\%$ of its original value for all
higher in-plane fields. The amplitude of $R_{yy}$, parallel to $B_{ip}$, rises
dramatically from zero over the same, initial
field range, reaches a value of about half of the initial $R_{xx}$ and increases
somewhat beyond this level for higher
in-plane fields. The amplitudes of $R_{xx}$ and $R_{yy}$ obviously trade places in
Fig.~3c, while $R_{xx}$ always exceeds $R_{yy}$
in Fig.~3a. However, the initial 
behavior for $B_{ip} \lesssim$ 1 T is very similar for
$R_{xx}$ and $R_{yy}$ in both panels and the
behavior is similar again for $B_{ip} \gtrsim$ 2 T 
albeit $R_{xx}$ and $R_{yy}$ having traded
places in Fig.~3c in the interim regime.
In fact, disregarding the narrow field region of the minimum in panel a)
and crossing in panel b), the general
pattern exhibited by the data of both panels is remarkably similar. 

For $\nu$ = 5/2 in Fig.~3b and 3d , the amplitudes of $R_{xx}$ and $R_{yy}$ are
essentially identical for $B_{ip}$ = 0 but they
separate as $B_{ip}$ increases. The order of $R_{xx}$ and $R_{yy}$ reflects the order of
$R_{xx}$ and $R_{yy}$ in the high-field region of
$\nu$ = 9/2 in Fig.~3a and 3c. However, the separation of 
$R_{xx}$ from $R_{yy}$ is gradual
and lacks any sharp transition
regime.

The top four panels of Fig.~3 are further summarized in the bottom
panels, which show the in-plane
anisotropy parameter for $\nu$ = 9/2 (Fig.~3e) and for $\nu$ = 5/2 (Fig.~3f) as a
function of $B_{ip}$ for both in-plane directions.
We define the anisotropy parameter as the ratio of the difference in
amplitudes divided by their sum. The solid
circles refers to $B_{ip}$ along the x-direction, the open circles refers to 
$B_{ip}$
along the y-direction, as depicted by the
abbreviated insets next to the traces. This panel shows very clearly that
for $\nu$ = 9/2 an in-plane magnetic field
along the originally hard direction, x, largely {\it preserves} the directional
anisotropy, whereas an in-plane magnetic
field along the originally easy direction, y, {\it reverses} the direction of
anisotropy. An in-plane field of $B{ip} \sim$ 1 T - 2 T
is sufficient to invert the anisotropy , {\it i.e.} rotate the underlying
electronic state by $\sim 90^{\circ}$ in the plane. The $\nu$ = 5/2
state, on the other hand, starts out isotropic and gradually develops an
anisotropy whose directionality at large
$B_{ip}$ is similar in extend to the one of $\nu$ = 9/2 in the neighboring panel.

At present the nature of the state at $\nu$ = 9/2 (as well as 11/2, 13/2,
15/2, $\cdots$) remains unresolved. Electronic states
akin to a charge density wave \cite{koulakov:prl96,moessner:prb96} 
or a liquid crystal state \cite{fradkin:cont9810148} are being
proposed, which would give rise to
anisotropic transport in the plane of the 2DES. Earlier experiments on the
anisotropy of $R$ in perpendicular
magnetic field found the hard and easy direction of transport to be pinned
to the lattice of the sample. This
spontaneous symmetry breaking is conjectured to arise from a slight
misalignment of the GaAs/AlGaAs
interface with respect to the [100] direction of the crystal, which causes
mono-atomic steps at the interface in a
particular direction within the plane \cite{lilly:prl99,du:ssc99}. 
The electronic phase at
half-filling in these higher Landau levels is
believed to align itself with respect to these steps, leading to
anisotropic transport in a given direction.

Our data in tilted magnetic fields indicate that this initial pinning
of the anisotropic electronic phase can be
overcome by an in-plane magnetic field of $B_{ip} \sim$ 1 T - 2 T. For fields higher
than this value the directionality of
the anisotropic phase is governed by the direction of the in-plane magnetic
field. A resistance measurement
with the current flowing parallel to this in-plane field always generates a
maximum in $R$ at $\nu$ = 9/2 and equivalent
states, whereas such a measurement performed with the current flowing
perpendicular to the in-plane field
generates a minimum in $R$ at $\nu$ = 9/2 and equivalent states. The actual angular
dependencies of the amplitudes of
$R$ (Fig.~3a,c) are rather non-monotonic and worth noting. 

Let us assume a simple model of an electrically anisotropic phase with
a preferential direction initially along
one of the sample edges and eventually along the in-plane field. In Fig.~3c
these directions are at 90$^{\circ}$ with
respect to each other and the phase flips orientation over a narrow field
range. In Fig.~3a these directions are
parallel to each other and no flip occurs, since the phase is already
aligned in the favorable direction. Under such
conditions one would assume $R$ to be largely angular independent, or at most
to vary gradually with angle. Yet
$R_{xx}$ drops rapidly, by a factor of two, for small 
$B_{ip}$ in Fig.~3a. A slight
misalignment of the initial phase with
respect to the x-axis is not expected to have such a dramatic effect. We
conclude that an explanation of the
angular dependencies of $R$ in Fig.~3a and Fig.~3c requires a more complex
model than a rigidly and smoothly
rotating electronic phase.

While the nature of the state at $\nu$ = 5/2 also remains obscure, it is
believed to be quite distinct from the state at
9/2. For one, the former is a true FQHE state \cite{willett:prl87} with plateau formation in
$R_{xy}$, whereas such a plateau seems to be
absent for the latter. And secondly, dramatic anisotropies in electronic
transport in purely perpendicular
magnetic field were only observed for the states at 9/2 and equivalent,
whereas they were absent in the 5/2 state.
Our tilted field experiments demonstrate that anisotropies not unlike those
of the 9/2 state can be induced in the
5/2 state at sufficiently high in-plane magnetic field. On the other hand,
such anisotropies have not been
observed for the states at $\nu$ = 3/2 and $\nu$ = 1/2.

The mechanism for the drastic influence of an in-plane magnetic field on
the transport properties of the 2D
system is unresolved. Any non-zero in-plane magnetic field increases the
total magnetic field, $B$, and therefore
increases the Zeeman energy, $E_Z$, since the spin experiences the full $B$ and
not only its perpendicular component.
Such a variation in $E_Z$ is known to cause angular dependent coincidences
between orbital energy and spin energy
leading to the disappearance and reappearance of energy gaps in the IQHE
and FQHE and hence to a strong
angular dependence of $R$. In fact, the earlier observed disappearance of the
FQHE at $\nu$ = 5/2 under tilted magnetic
field was considered to be of such a spin origin. However, no spin
mechanism has been brought forward that
would create a macroscopic electrical anisotropy, such as in the 5/2-state
under tilt, or provide a preferred
direction for an existing anisotropic state such as at $\nu$ = 9/2. At present we
are not aware of a model for our
observations.

We would like to thank E. Palm and T. Murphy for experimental assistance,
and B. I. Altshuler and S. H. Simon
for discussions. A portion of this work was performed at the National High
Magnetic Field Laboratory which is
supported by NSF Cooperative Agreement No. DMR-9527035 and by the State of
Florida. D.C.T. and W.P. are
supported by NSF and by the DOE. R.R. Du is supported by NSF and by the
Sloan Foundation.

\begin{figure}
\vspace{0.2cm}
\centerline{\psfig{figure=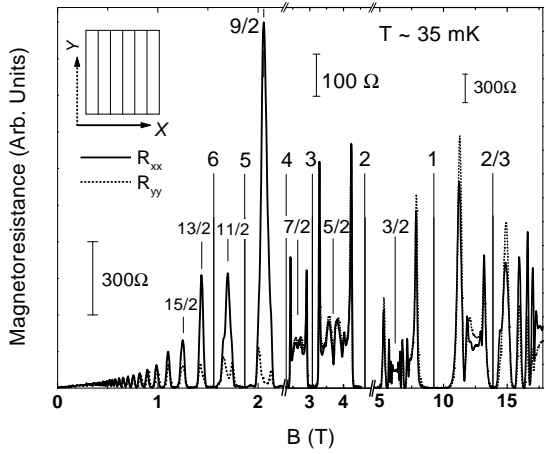,width=7.5cm,angle=0}}
\vspace{0.2cm}
\caption{
Overview of magneto resistance of our high-mobility sample in perpendicular
magnetic field. The
features of the IQHE ($\nu$ = 1, 2, 3, $\cdots$) and 
FQHE ($\nu$ = 2/3 etc) are clearly
visible. Positions of half-filling are marked
from $\nu$ = 3/2 to $\nu$ = 15/2. The insert 
shows the sample and the directions x and
y. The magneto resistances $R_{xx}$ and
$R_{yy}$, taken in the x-direction and 
y-direction, respectively, in the plane
are very similar except around
half-filling of higher Landau levels ($\nu$ = 9/2 to $\nu$ = 15/2) 
where they strongly
differ.
}
\vspace{0.2cm}
\end{figure}

\begin{figure}
\vspace{0.2cm}
\centerline{\psfig{figure=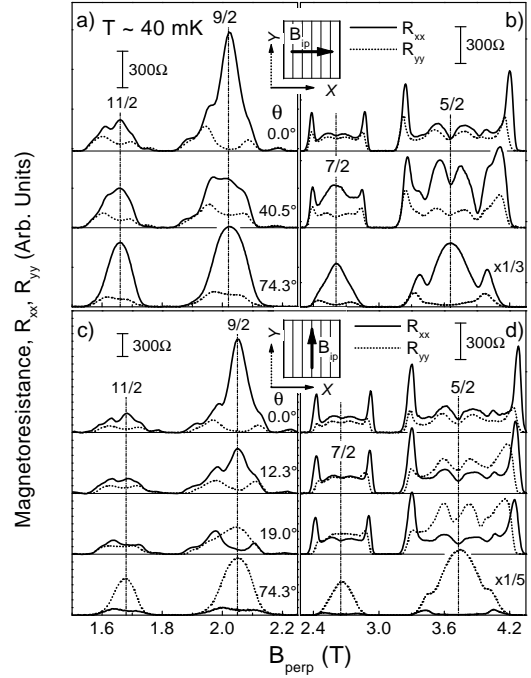,width=7.5cm,angle=0}}
\vspace{0.2cm}
\caption{
Dependence of the magneto resistance 
$R_{xx}$ and $R_{yy}$ around filling factor 9/2
and 11/2 as well as around
5/2 and 7/2 on angle, $\theta$, and direction of 
a tilted magnetic field, $B$. $B_{perp}$
represents the field perpendicular to the
sample, $B_{perp} = B \times$ cos($\theta$). 
The sample geometries are depicted as inserts.
The x and y-directions are fixed with
respect to the sample. Stripes in the sample indicate the initial
anisotropy of the 9/2 and 7/2 state. In panel  a)
and b) the sample is rotated around the y-axis generating an increasing
in-plane field $B_{ip} = B \times$ sin($\theta$) along the
hard direction, x, whereas in panel c) and d) the sample is rotated around
the x-axis generating an increasing
$B_{ip}$ along the easy direction, y.
}
\vspace{0.2cm}
\end{figure}

\begin{figure}
\vspace{0.2cm}
\centerline{\psfig{figure=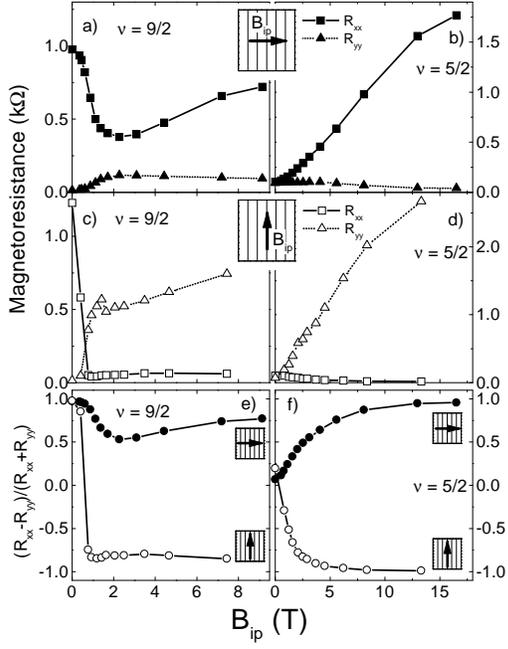,width=7.5cm,angle=0}}
\vspace{0.2cm}
\caption{
Amplitudes of $R_{xx}$ and $R_{yy}$ at 
$\nu$ = 9/2 and $\nu$ = 5/2 as a function of in-plane
magnetic field $B_{ip}$. The top four
panels match the four panels of Fig.~2. Inserts depict the sample
geometries. The bottom panels show the
anisotropy factor determined from the amplitudes of the panels above,
separate for the 9/2 state (panel e)) and
5/2 state (panel f)).
}
\vspace{0.2cm}
\end{figure}

\end{document}